\documentstyle[12pt]{article}
\begin{document}
\topmargin -1.4truecm
\textheight 8.8 in
\title{ \hfill
hep-ph/0006138\\ \vskip 1.3truecm {\large{\bf $U(1)$ symmetry and $R$ parity 
violation}}} \vskip 1.5truecm

\author{ Anjan S. Joshipura, Rishikesh Vaidya and Sudhir K. Vempati \\
{\ns\it Theoretical Physics Group, Physical Research Laboratory,}\\
{\ns\it Navarangpura, Ahmedabad, 380 009, India.}}
\date{}
%------------------------------------------------------------
\def\ap#1#2#3{           {\it Ann. Phys. (NY) }{\bf #1} (19#2) #3}
\def\arnps#1#2#3{        {\it Ann. Rev. Nucl. Part. Sci. }{\bf #1} (19#2) #3}
\def\cnpp#1#2#3{        {\it Comm. Nucl. Part. Phys. }{\bf #1} (19#2) #3}
\def\apj#1#2#3{          {\it Astrophys. J. }{\bf #1} (19#2) #3}
\def\asr#1#2#3{          {\it Astrophys. Space Rev. }{\bf #1} (19#2) #3}
\def\ass#1#2#3{          {\it Astrophys. Space Sci. }{\bf #1} (19#2) #3}

\def\apjl#1#2#3{         {\it Astrophys. J. Lett. }{\bf #1} (19#2) #3}
\def\ass#1#2#3{          {\it Astrophys. Space Sci. }{\bf #1} (19#2) #3}
\def\jel#1#2#3{         {\it Journal Europhys. Lett. }{\bf #1} (19#2) #3}

\def\ib#1#2#3{           {\it ibid. }{\bf #1} (19#2) #3}
\def\nat#1#2#3{          {\it Nature }{\bf #1} (19#2) #3}
\def\nps#1#2#3{          {\it Nucl. Phys. B (Proc. Suppl.) }
                         {\bf #1} (19#2) #3}
\def\np#1#2#3{           {\it Nucl. Phys. }{\bf #1} (19#2) #3}
\def\pl#1#2#3{           {\it Phys. Lett. }{\bf #1} (19#2) #3}
\def\pr#1#2#3{           {\it Phys. Rev. }{\bf #1} (19#2) #3}
\def\prep#1#2#3{         {\it Phys. Rep. }{\bf #1} (19#2) #3}
\def\prl#1#2#3{          {\it Phys. Rev. Lett. }{\bf #1} (19#2) #3}
\def\pw#1#2#3{          {\it Particle World }{\bf #1} (19#2) #3}
\def\ptp#1#2#3{          {\it Prog. Theor. Phys. }{\bf #1} (19#2) #3}
\def\jppnp#1#2#3{         {\it J. Prog. Part. Nucl. Phys. }{\bf #1} (19#2) #3}

\def\rpp#1#2#3{         {\it Rep. on Prog. in Phys. }{\bf #1} (19#2) #3}
\def\ptps#1#2#3{         {\it Prog. Theor. Phys. Suppl. }{\bf #1} (19#2) #3}
\def\rmp#1#2#3{          {\it Rev. Mod. Phys. }{\bf #1} (19#2) #3}
\def\zp#1#2#3{           {\it Zeit. fur Physik }{\bf #1} (19#2) #3}
\def\fp#1#2#3{           {\it Fortschr. Phys. }{\bf #1} (19#2) #3}
\def\Zp#1#2#3{           {\it Z. Physik }{\bf #1} (19#2) #3}
\def\Sci#1#2#3{          {\it Science }{\bf #1} (19#2) #3}
\def\n.c.#1#2#3{         {\it Nuovo Cim. }{\bf #1} (19#2) #3}
\def\r.n.c.#1#2#3{       {\it Riv. del Nuovo Cim. }{\bf #1} (19#2) #3}
\def\sjnp#1#2#3{         {\it Sov. J. Nucl. Phys. }{\bf #1} (19#2) #3}
\def\yf#1#2#3{           {\it Yad. Fiz. }{\bf #1} (19#2) #3}
\def\zetf#1#2#3{         {\it Z. Eksp. Teor. Fiz. }{\bf #1} (19#2) #3}
\def\zetfpr#1#2#3{         {\it Z. Eksp. Teor. Fiz. Pisma. Red. }{\bf #1} (19#2) #3}
\def\jetp#1#2#3{         {\it JETP }{\bf #1} (19#2) #3}
\def\mpl#1#2#3{          {\it Mod. Phys. Lett. }{\bf #1} (19#2) #3}
\def\ufn#1#2#3{          {\it Usp. Fiz. Naut. }{\bf #1} (19#2) #3}
\def\sp#1#2#3{           {\it Sov. Phys.-Usp.}{\bf #1} (19#2) #3}
\def\ppnp#1#2#3{           {\it Prog. Part. Nucl. Phys. }{\bf #1} (19#2) #3}
\def\cnpp#1#2#3{           {\it Comm. Nucl. Part. Phys. }{\bf #1} (19#2) #3}
\def\ijmp#1#2#3{           {\it Int. J. Mod. Phys. }{\bf #1} (19#2) #3}
\def\ic#1#2#3{           {\it Investigaci\'on y Ciencia }{\bf #1} (19#2) #3}
\def\tp{these proceedings}
\def\pc{private communication}
\def\ip{in preparation}
\relax
\newcommand{\GeV}{\,{\rm GeV}}
\newcommand{\MeV}{\,{\rm MeV}}
\newcommand{\keV}{\,{\rm keV}}
\newcommand{\eV}{\,{\rm eV}}
\newcommand{\Tr}{{\rm Tr}\!}
\renewcommand{\arraystretch}{1.2}
\newcommand{\beq}{\begin{equation}}
\newcommand{\eeq}{\end{equation}}
\newcommand{\beqa}{\begin{eqnarray}}
\newcommand{\eeqa}{\end{eqnarray}}
\newcommand{\ba}{\begin{array}}
\newcommand{\ea}{\end{array}}
\newcommand{\bmat}{\left(\ba}
\newcommand{\emat}{\ea\right)}
\newcommand{\refs}[1]{(\ref{#1})}
\newcommand{\ler}{\stackrel{\scriptstyle <}{\scriptstyle\sim}}
\newcommand{\ger}{\stackrel{\scriptstyle >}{\scriptstyle\sim}}
\newcommand{\lag}{\langle}
\newcommand{\rag}{\rangle}
\newcommand{\ns}{\normalsize}
\newcommand{\cm}{{\cal M}}
\newcommand{\gr}{m_{3/2}}
\newcommand{\p}{\partial}
\setcounter{page}{0}
\def\u1x{ $U(1)$}
\def\rp{ $R_P$}
\def\321{$SU(3)\times SU(2)\times U(1)$}
\def\tl{{\tilde{l}}}
\def\tL{{\tilde{L}}}
\def\bd{{\overline{d}}}
\def\tL{{\tilde{L}}}
\def\a{\alpha}
\def\b{\beta}
\def\g{\gamma}
\def\c{\chi}
\def\d{\delta}
\def\D{\Delta}
\def\db{{\overline{\delta}}}
\def\Db{{\overline{\Delta}}}
\def\e{\epsilon}
\def\l{\lambda}
\def\n{\nu}
\def\m{\mu}
\def\nt{{\tilde{\nu}}}
\def\p{\phi}
\def\P{\Phi}
\def\x{\xi}
\def\r{\rho}
\def\s{\sigma}
\def\t{\tau}
\def\th{\theta}
\def\ne{\nu_e}
\def\nm{\nu_{\mu}}
\def\rp{$R_P$}
\def\mp{$M_P$}
\def\emt{$L_e-L_{\mu}-L_{\tau}$}
\renewcommand{\Huge}{\Large}
\renewcommand{\LARGE}{\Large}
\renewcommand{\Large}{\large}
\maketitle
\thispagestyle{empty}
\vskip 1.5truecm
\begin{center}
\underline{\bf{ABSTRACT}}
\end{center}
\vskip 0.8truecm
\begin{abstract}
The patterns of $R$ violation resulting from imposition of a gauged
$U(1)$ horizontal symmetry on the minimal supersymmetric standard model
are
systematically analyzed. We concentrate on  class of models
with integer $U(1)$ charges chosen to reproduce the quark masses and
mixings as
well as charged lepton masses exactly or approximately.
The $U(1)$ charges are further restricted by the requirement that
very large bilinear lepton number violating terms should not be allowed
in the superpotential. It is shown that this leads to severely
constrained patterns of trilinear interactions. Specifically, only choice
compatible with phenomenological restrictions is the one in which all  
the trilinear $\lambda'_{ijk}$ and all but at most two trilinear
$\lambda_{ijk}$ couplings vanish or are enormously suppressed. The
$U(1)$ symmetry can allow effective generation of bilinear lepton number
violating parameters through  terms in the Kahler
potential. Resulting models are identified and structure of
neutrino masses in some of these is briefly discussed.
\end{abstract} 
\newpage
\section{Introduction}

One of the attractive ways to understand the mysterious  hierarchy among
quark and lepton masses is to postulate the existence of a 
U(1) symmetry broken spontaneously at a scale much larger than that of
weak interactions \cite{FN}. Most fermion masses and the entire Cabibbo Kobayashi Maskawa
(CKM) matrix arise in this approach due to breaking of the \u1x and are determined in 
terms of a parameter $\lambda\sim {<\theta>\over M}$ and the \u1x
charges of the fermions. Here $<\theta>$ determines the  scale of \u1x breaking and $M$ is 
some higher scale which could be the Planck scale $M_P$ or the string scale if
\u1x arises from an underlying string theory. The $\l$ is usually identified with the 
Cabibbo angle $\sim 0.22$ and all the fermion mass matrices are represented
as powers of $\l$. 
Although this mechanism is quite general, it becomes quite attractive
to combine the virtues of this \u1x symmetry with that of the minimal supersymmetric
standard model (MSSM) \cite{seiberg,ibanez,ramond,dudas1,ramond2,chun,rest}. 
In this case, the \u1x can give information not only on the
quark spectrum but also on the $R$ parity violating couplings which can determine
the neutrino masses through the pattern of the R violation it dictates 
\cite{ramond2,chun,dudas2,ellis,rpreview}.
%Theoretically, both
%can  arise from a superstring theory. In this context,
%the \u1x symmetry needs to be a gauge symmetry and cancellation
%of the gauge anomalies then imposes further restrictions on possible charges carried by the 
%MSSM fields. Another advantage of using underlying supersymmetric framework
%follows from the automatic presence of the lepton number violating couplings in the 
%MSSM. These couplings can lead to neutrino masses and imposition of \u1x symmetry
%then can  lead to restrictions on the allowed forms of the neutrino mass matrix.

The lepton number violation in the MSSM is generated due to the presence of the 
supersymmetric partners of 
quarks and leptons. This can be characterized by the following $R$ violating terms in
the superpotential of the model:
\beq 
\label{super}
W_{{\not R}_p}= \l^{'}_{ijk}L_{i}Q_{j}D^{c}_{k} + \l_{ijk} L_{i}L_{j}E^{c}_{k} + \e_{i} L_{i}H_{2}
\eeq
A priori, this involves 39 independent parameters. Each of this can contribute 
to the mass matrix for the 
three light neutrinos. It is desirable to restrict the number of the allowed couplings
from some symmetry 
principle and the \u1x symmetry can play a crucial role. By requiring that the \u1x
charges of the MSSM field should be such that it leads to correct quark and charge 
lepton masses as well as  
the CKM matrix, one could considerably reduce the freedom in choosing the \u1x charges.
Set of charges so determined would lead to definite  patterns of the $R$
violating couplings appearing in eq.(\ref{super}). This in turn leads to specific structure for 
neutrino masses. 
%There have been number of papers which looked at the structure
%of the $R$ violating couplings in the presence of a \u1x.
%It is not guaranteed  that the pattern of the $R$ violating couplings predicted 
%this way 
%will be consistent with the stringent phenomenological constraints imposed on them from
%different processes notably from the flavour violating transitions. 

The purpose of this note is to systematically search for all possible allowed 
patterns for the $R$  violating couplings of eq.(\ref{super}) which  result from  \u1x
charge assignments consistent with the successful predictions in the quark sector 
in case of the integer \u1x charges for all the fields. In a large class of such
models \cite{ramond,chun,ellis,rpreview}, the  \u1x symmetry tends to lead to very large 
and phenomenologically unacceptable values for the 
coefficient $\epsilon_i$ of the bilinear terms in eq.(\ref{super}). Requiring that this does 
not happen restricts the allowed set of models in a stringent manner. We find 
a remarkable result that in all these restricted models,
almost all the trilinear couplings in eq.(\ref{super}) are either zero, highly suppressed or
 their predicted magnitudes
are inconsistent with phenomenology. Specifically, all the models we analyzed
require zero $\lambda'_{ijk}$  and at most one  or two non-zero $\lambda_{ijk}$ 
if they are to be phenomenologically 
consistent. The resulting theory still possesses  lepton number 
violation since significant amount of  bilinear couplings can be generated
through couplings in Kahler potential using the mechanism proposed by
Giudice and Masiero \cite{GM}. The neutrino mass patterns in this case gets restricted 
in terms of only three or four 
independent lepton number violating parameters making  \u1x symmetry very 
predictive scheme not only for the 
descriptions of  the quark spectrum but also for the neutrino masses and mixing.

We start in the next section  with a discussion of our  framework and  the basic
assumptions and highlight the problem of generation of the large $\epsilon_i$ 
parameters within this framework. In the next section, we discuss the structure
of trilinear interactions and their consistency with phenomenology in models
which can explain the quark spectrum. Section (4) contains specific discussion of
the consequences of models allowed on phenomenological ground and we summarize 
the main results in the last section. 

\section{\u1x symmetry and $\epsilon$ problem} 

Let us consider the MSSM augmented with a gauged horizontal \u1x symmetry.
The standard superfields $(L_i,~ Q_i,~ D_i^c,~ U_i^c,~ E_i^c,~ H_1,~ H_2)$
are assumed to carry the charges $(l_i,~ q_i,~ d_i,~ u_i,~ e_i,~ h_1,~ h_2)$
respectively with i running from 1 to 3. The \u1x symmetry is assumed to be broken at a high scale by
the vacuum expectation value (VEV) of one gauge singlet superfield $\theta$
with the \u1x charge normalized to -1 or with two such fields $\theta,\bar{\theta}$ 
with charges -1 and 1 respectively. It is normally assumed that only 
the third generation of fermions have renormalizable couplings invariant
under \u1x. The rest of the couplings arise in the effective theory from 
higher dimensional terms \cite{FN}:

$$ \Psi_i \Psi_j H \left( {\th \over M} \right)^{n_{ij}}$$
where $\Psi_i$ is a chiral superfield, H is the Higgs doublet and M is some
higher mass scale which could be the Planck scale $M_p$ and $n_{ij} =
\psi_i + \psi_j$ are positive numbers representing the charges of
$\Psi_i$, $\Psi_j$ under \u1x respectively. Similar term is absent in case of
a negative $n_{ij}$ due to holomorphic nature of W \cite{seiberg}.
 For positive $n_{ij}$, one gets an 
 $ij^{th}$ entry of order $\lambda^{n_{ij}}$ in the mass matrix for the field $\Psi$. 
Identification $\lambda\sim 0.22$ and proper choice of the \u1x charges leads to successful quark
mass matrices \cite{ibanez,ramond,dudas1}.
 
A priori, the model has 15 independent  \u1x charges for matter and 2
charges for Higgs fields. Of these, all but four can be fixed  from 
different requirements discussed in the literature which we list
below \cite{dudas1}. \\

\noindent (1) The fermions in the third generation are  assumed to have the following
couplings invariant under \u1x
\beq
W_Y = \b_t Q_3 U_3^c H_2  + \b_b Q_3 D_3^c H_1 \left( \th \over M \right)^x + 
\b_\tau L_3 E_3^c H_1 \left( \th \over M \right)^x
\eeq
This is possible if,
\beq
\label{renorm}
\ba{lll}
q_{3} + u_{3} + h_{2} = 0 ; \;\;\;& q_{3} + d_{3} + h_{1} = & l_{3} + e_{3} + h_{1} = x
\ea
\eeq
This determines $h_2=-q_3-u_3$ and $h_1=-q_3-d_3+x$ with $\tan\b\sim \lambda^x (m_t/m_b)$.
The phenomenological requirement of $\tan\beta\geq O(1)$ implies $0\leq x\leq 2$.\\
$b - \tau$ unification has been implicitly assumed in writing eq.(\ref{renorm}).\\

\noindent (2) The charge differences $q_{i3}\equiv q_i-q_3,~ u_{i3}\equiv u_i-u_3$ 
and $d_{i3}\equiv d_i-d_3$ ($i=1,2$) are determined
by requiring that the quark masses and the CKM matrix come out to be exactly or
approximately correct. Various possible values for these differences have been 
classified in  \cite{dudas1} and we shall use these results.\\

\noindent (3) The \u1x symmetry being gauged is required to be anomaly free. It has been shown 
 \cite{ramond} that all the relevant \u1x anomalies  cannot be zero in models with a single
$\theta$ if one is to require the correct structure for the quark and lepton masses.
These anomalies then needs to be cancelled by the Green-Schwarz mechanism \cite{GS}.
This requirement imposes three non-trivial relations among the \u1x charges.\\

\noindent (4) The prediction of approximately correct hierarchy among the 
charged lepton masses requires 

\beq
\label{clepton}
\ba{ll}
l_{13} + e_{13} = 4~ \mbox{OR}~ 5 \;\; ; \;\; & l_{23} + e_{23} = 2
\ea
\eeq

After imposing the above listed requirements, the successful model is fixed in terms
of the 4 independent charges. Each choice of these charges  would imply 
different patterns for $R$ violation. Since the \u1x is capable of predicting
orders of magnitudes of various couplings, it is not guaranteed that all the 
patterns of R violation predicted in this way would be phenomenologically 
consistent. In fact very few can meet the constraints from phenomenology. The most 
stringent constraint on possible choice of $R$ charges is provided by 
the parameters $\e_i$. The \u1x symmetry can lead to the following term in
W:
\beq M\;L_i\;H_2 \left({\theta\over M}\right)^{l_i+h_2} \eeq
This leads to 
\beq 
\label{eps}
\e_i\sim M \left({<\theta>\over M}\right)^{l_i+h_2} \sim M \lambda^{l_i+h_2}
\eeq
Unless the charges $l_i+h_2$ are appropriately chosen, the predicted value
for $\e_i$ can grossly conflict with (a) the scale of $SU(2) \times U(1)$ breaking
which would require sneutrino VEV $\leq  O(M_W)$ and (b) neutrino masses.
A bilinear parameter $\e$ would imply a neutrino mass \cite{hs} of order \cite{asjmarek}:
\beq
\label{numass}
m_\n\sim \left({\e\over \m}\right)^2 {M_Z^2\over M_{SUSY}}\sin^2 \phi \eeq
Here, $\sin^2 \phi$ is O(1) if SUSY breaking is not characterized by
the universal boundary conditions at a high scale. In the converse case,
this factor gets enormously suppressed due to the fact that $\e_i$ can be rotated
away from the full Lagrangian in the limit of vanishing down quark and 
charged lepton couplings. This issue is discussed in number of papers \cite{bilinears}.
 Typical order of magnitude estimate of $\sin^2\phi$ is \cite{asjbabu}
 
\beq 
\sin^2 \phi \sim \left({3 h_b^2 ln{m_X^2\over m_Z^2}\over 16 \pi^2}\right)^2
\sim 10^{-7} 
\eeq

These equations are very rough estimates. The exact values depend upon the MSSM parameters.
But these rough estimates are sufficient to show that phenomenologically required 
$\e_i$ are grossly in disagreement with the typical predictions, for e.g, even with
$sin \phi^2 \sim 10^{-7}$, $m_{\nu} <  eV$ would need $\e \sim$ GeV for $~\mu \sim
M_{SUSY} \sim 100$ GeV.  

In order to prevent very  large $\e_i$ being generated, 
one must ensure one  of the following:\\

\noindent (a) $l_i+h_2$ is bounded by  
\beq
\label{bound}
l_i+h_2 \ger 24. 
\eeq
This can lead to $\e_i$ in  $\GeV$  range and neutrinos with mass in the $\eV$ range
in case of models with universal boundary conditions and $M\sim
10^{16}\GeV$.  In models without the universal boundary conditions, 
the required magnitude for $l_i+h_2$ would be even larger.\\

\noindent
(b) \u1x is broken by only one superfield $\theta$ and $l_i+h_2$ is 
negative. The terms in eq.(\ref{eps}) are then not allowed in $W$ 
by the \u1x symmetry and by the analyticity of W.\\

\noindent
(c)  $l_i+h_2$ is fractional, forbidding coupling of bilinear term to $\theta$.\\

\noindent
(d) Impose some additional symmetry, e.g. modular invariance which may prevent
occurrence of dangerous terms \cite{dudas3}.
 
Note that  models containing two $\theta$-like fields with opposite \u1x
charges would lead to large $\e_i$ independent of the sign of $l_i+h_2$. 
Thus these models
can be made phenomenologically consistent only by choosing fractional or 
unnaturally high values for $|l_i+h_2|$. We shall therefore not consider these models
and concentrate only on models with a single $\theta$ and also assume
only integer \u1x charges. Then $\e_i$ can be suppressed either through (a) 
or through (b) if no other symmetry is imposed.
 
Although the structure of $R$ violating interactions following from a \u1x 
symmetry alone has been discussed in a number of papers \cite{ramond,chun,dudas2,ellis,rpreview},
 the requirement that the \u1x symmetry should not generate large $\e_i$
 has not always been imposed \cite{ramond,chun,ellis}.
It is argued customarily that  $\epsilon_i$
are unphysical as they can be rotated away by redefining the new $H_1$ as a linear   
combination of the original $H_1$ and $L_i$ appearing in eq.(\ref{super}). This however
changes the original $\mu$ parameter to $(\mu^2+\e_i^2)^{1/2}$. Thus if 
the models do allow large $\e_i$ then rotating them away generates equally large
$\mu$ which is also phenomenologically inconsistent. One must therefore allow only 
the \u1x charge assignments corresponding to zero or suppressed $\e_i$ in $W$.
\section{Structures of trilinear couplings}
In this section, we shall enumerate possible \u1x models leading to correct
quark mass spectrum and investigate structures for the 
trilinear couplings in these models keeping the phenomenological constraints 
in mind.

After imposing eqs.(\ref{renorm}), the quark mass ratios and the CKM mixing angles 
are determined in terms of the quark charge differences. Systematic search for the 
possible charge differences led to the eight models \cite{dudas1,chun} 
reproduced in the table below:\\

%\newpage
\vskip 0.5cm
\begin{center}
{\large Models}\\[20pt]
\begin{tabular}{|c|c|c|c|c|c|c|c|c|} \hline
{\bf Models} & {$\bf l_{13}+e_{13}$} & {$\bf l_{23}+e_{23}$} & {$\bf
  q_{13}$} & {$\bf q_{23}$} & {$\bf u_{13}$} & {$\bf u_{23}$} & {$\bf d_{13}$}
& {$\bf d_{23}$} \\ \hline\hline

{\rm IA} & 4 & 2 & 3 & 2  & 5 & 2 & 1 & 0 \\  \hline
{\rm IIA} & 4 & 2 & 4 & 3 & 4 & 1 & 1 & -1 \\ \hline
{\rm IIIA} & 4 & 2 & 4 & 3& 4& 1& -1 & -1 \\ \hline
{\rm IVA} & 4 & 2 & -2 & -3 & 10 & 7 & 6 & 5 \\ \hline\hline
{\rm IB} & 5 & 2 & 3 & 2 & 5 & 2 & 1 & 0 \\ \hline
{\rm IIB} & 5 & 2 & 4 & 3 & 4 & 1 & 1 & -1 \\ \hline
{\rm IIIB} & 5 & 2 & 4 & 3 & 4 & 1 & -1 & -1 \\ \hline
{\rm IVB} & 5 & 2 & -2 & -3 & 10 & 7 & 6 & 5 \\ \hline
\end{tabular}
\end{center}
%\begin{center}
{\bf Table 1 :} We present here all the possible models which generate correct
quark and lepton mass hierarchies as well as the CKM matrix. 
%\end{center}

%beq 
%label{models}
%ba{llllllll}
%_{13} = 3; & q_{23}  = 2; & u_{13}  =  5;  & u_{23}  = 2;  & d_{13}
%1; & d_{23}  =  0 && {\rm I}  \\
%_{13} = 4; & q_{23} = 3; & u_{13} = 4; & u_{23}  =  1; &  d_{13}  =
%; & d_{23} = -1 & & {\rm II} \\
%_{13} = 4; & q_{23} = 2; & u_{13} = 5; & u_{23} = 2; & d_{13} = -1; &
%_{23} = -1 & & {\rm III} \\
%_{13}=-2; & q_{23} =-3; & u_{13}=10; & u_{23} =7; & d_{13} = 6; &
%_{23} = 5 & & {\rm IV}
%ea \eeq
\vskip 0.5cm

The model {\rm I} exactly reproduces the quark mass ratios and all the three
CKM mixing angles. Since the predictions of the \u1x symmetry are exact only up to 
coefficients of O(1), one has to allow for models which may deviate from the exact
predictions by small amount. The charge differences in model {\rm
  II}, {\rm III}, and {\rm IV}  represent the models
which deviate from the exact predictions by O($\lambda$) \cite{dudas1}. 
The leptonic mixing 
analogous to the CKM matrix is still arbitrary in these models but the charged 
lepton masses are required to satisfy ${m_e\over m_{\t}}\sim \lambda^4,{m_\m\over 
m_{\t}}\sim \lambda^2$ in models (A) and ${m_e\over m_{\t}}\sim \lambda^5,{m_\m\over
m_\t}\sim \lambda^2$ in models (B).

The \u1x charges are still subject to the anomaly constraint.
The anomalies generated due to the presence of the extra \u1x are as follows:

\beqa
\label{anom}
[SU(3)]^{2}U(1)_{X}:~~~  
        A_{3} &=& \sum_{i=1}^{3} (2q_{i} +u_{i} + d_{i}) \nonumber  \\[1.5mm] 
[SU(2)]^{2}U(1)_{X}:~~~  
        A_{2} &=& \sum_{i=1}^{3} (3q_{i} +l_{i}) + h_{1} + h_{2} \nonumber \\[1.5mm]
[U(1)_{Y}]^{2}U(1)_{X}:~~~ 
  A_{1} &=& \sum_{i=1}^{3} ({1\over 3}q_{i} + {8\over 3} u_{i} + {2\over 3} d_{i} + l_{i} + 2e_{i}) + h_{1}
+h_{2}  \nonumber \\[1.5mm]
U(1)_{Y}[U(1)]_{X}^{2}:~~~
 A'_{1} &=& \sum_{i=1}^{3}(q_{i}^{2} -2u_{i}^{2} + d_{i}^{2} - l_{i}^{2} + e_{i}^{2}) -h_{1}^{2} + h_{2}^{2}
\eeqa
These can be cancelled in string theory through the Green-Schwartz mechanism \cite{GS}
by requiring 
\beq
\label{gs}
A_{2}=A_{3}={3\over 5}A_{1}; ~~~A'_{1} = 0.
\eeq 
The above constraints on $A_1, A_2, A_3$ can be solved to give:
\beqa
\label{h}
h\equiv h_1+h_2&=&\sum_{i=1}^{3} \left( q_{i3} + d_{i3} \right)  - 
\sum _{i=1}^3 \left( l_{i3} + e_{i3} \right) ,\nonumber \\
l_2 &=& m -(l_{1} + l_{3} + 9q_{3} + 4h -3x),
\eeqa
where 
\beq
\label{em}
m = \sum_{i=1}^{3}(u_{i3} + d_{i3} - q_{i3}). 
\eeq
Also from eqs.(\ref{renorm}),
\beq
\label{u3}
u_3 = x - 2 q_3 - d_3 - h
\eeq 
Note that the parameter h determines whether the $\mu$ term is allowed in $W$.
Positive $h$ will result in too large $\mu$ unless $h$ is also correspondingly large
\footnote{see however ref. \cite{dudas3} which imposes additional modular invariance}.
Negative $h$ does not allow the $\m$ term in $W$ but phenomenologically consistent value
can be generated through GM mechanism in this case. $h=0$ allows arbitrary $\mu$ in $W$.
The anomaly constraints determines $h$ completely in terms of the charge differences
fixed by the models in Table 1 and is insensitive to the overall redefinition of the
\u1x charges. It is seen that all except model ({\rm IIA}) lead to zero or negative 
$h$ and thus are phenomenologically consistent.

The magnitudes and structure of the trilinear couplings is determined by 
the following equation:
\beqa
\label{lamb}
\lambda'_{ijk}&=&\theta(c_i + n_{jk}^d)\lambda^{c_i + n_{jk}^d} \nonumber \\
\lambda_{ijk}&=& \theta(c_i + n_{jk}^l)\lambda^{c_i + n_{jk}^l} 
\eeqa
where $c_i=l_i+x+h_2-h$ ; $n^{d}_{jk}=q_{j3}+d_{k3}$ ;  $n^{l}_{jk} = l_{j3} + e_{k3}$ with
$n_{jk}^d, n_{jk}^l$ being completely fixed for a given model displayed in Table 1. 
Note that some of the trilinear couplings may be zero if the corresponding exponent is
negative. They may still be generated due to non-minimal contribution
to the kinetic energy term of different fields \cite{dudas1,ramond2,chun}. Such contributions do not however affect 
the order of magnitudes of those couplings which are non-zero to start with \cite{ramond2}.

After imposing the constraints of eqs.(\ref{gs}), one is still left with four
independent parameters including $x$. One would thus expect considerable freedom
in the choice of $\lambda'_{ijk},\lambda_{ijk}$. Typically, more than one such couplings 
are allowed to be non-zero simultaneously in various models. Thus they
lead to flavour violating transitions which are known to be enormously suppressed.
It is these constraints on the product of trilinear couplings which lead to
stringent restrictions on the allowed \u1x charges. It turns out that
constraint following from the $K^0-\bar{K}^0$ mass difference 
 alone is sufficient to rule out
the presence of non-zero trilinear couplings in most models. The  $K^0-\bar{K}^0$
mass difference constrains the product $\l'_{i12}\l'_{i21}$ to be $\leq 10^{-9}$
\cite{allanach} for the slepton masses of O(100 GeV). Allowing for some variation 
in these masses, we shall use the following conservative limit
\beq
\label{k0k0}
\l'_{i12}\l'_{i21}\leq \l^{12}\sim 1.3 \cdot 10^{-8} 
%\l'_{i13}\l'_{i31}&\leq& \l^{9}\sim 1.2 \cdot 10^{-6} 
\eeq
%These along with the constraint implied by $e_i$ turn out to be
%quite restrictive and does not allow any non-trivial trilinear couplings in any of 
%the models as we now show. 
We now analyze the magnitudes of the product in eq.(\ref{k0k0})
predicted by  models of Table 1, when one imposes the additional 
requirement that the $l_i+h_2$ is negative or has a  large value given in  
eq.(\ref{bound}). These requirements 
result in  zero or suppressed $\e_i$ respectively. But they would also lead to
zero or suppressed trilinear interactions as we now discuss. Let us consider these 
two cases separately.

\subsection{$l_i+h_2\ger 24$}
In this case, $\e_i$ are artificially forced to be small by choosing
very large value of $l_i+h_2$ as in eq.(\ref{bound}). But the large value of these charges
also results in the enormous suppression in the allowed magnitudes
of the trilinear couplings. This is easily seen from eqs.(\ref{lamb}). Since
$h$ is zero or negative for all the allowed models, and $x\leq 2$, it follows that
$$ c_i=l_i+h_2+x-h ~\geq~ l_i+h_2 ~\geq~ 22\; .$$
It follows from Table 1 that the $n_{jk}^{d,l}$ are positive or small negative numbers
in all the models. As a consequence, all the trilinear couplings 
are $\leq \l^{19}\sim 10^{-12}$ in this case. 
This value is too small to have any phenomenological 
consequence. 
\subsection {$l_i+h_2<0$}
We shall first show that the most preferred model {\rm IA} can be phenomenologically
consistent in this case only when all $\lambda'_{ijk}$ are zero and then 
generalize this result to 
other cases. The $\l'_{ijk}$ are explicitly given as follows in this model:

\beq 
\label{lpri}
\l'_{ijk} = \l^{l_i + h_2 + x} \left[
\ba{ccc}
\l^4 & \l^3 & \l^3 \\
\l^3 & \l^2 & \l^2 \\
\l & 1 & 1 
\ea
\right]
\eeq

where it is implicit that some element is zero if corresponding exponent
is negative \cite{seiberg}. The matrix in the above eq. (\ref{lpri}) 
coincides with $\e^{-x} \left( M_d \right)_{jk}$. Hence for negative
$l_i + h_2$, it follows that the $\l'_{ijk}$ is either larger than the
matrix element $(M_d)_{jk}$ or is zero for every $i$. In the former case, one
cannot easily meet the phenomenological requirement in eq.(\ref{k0k0}). 
Specifically, equation for the $c_i$ gets translated to
\beqa
\label{ci}
c_i\equiv l_i+h_2+x &<& -3\;\;\;\;\;\; {\mbox OR} \nonumber \\
&\geq& \;\;\;3 \eeqa
This condition ensures that  $\l'_{i12}\l'_{i21}$ either
 satisfies eq.(\ref{k0k0})
(when $c_i>3$ ) or is identically zero when $c_i<-3$. But $c_i\geq 3$
is untenable since $l_i+h_2\leq 0$ and $\tan\beta\sim \l^{x}(m_t/m_b) \geq O(1)$ needs
 $x\leq 2$ leading to $c_i\leq 2$. As a result one must restrict  $c_i$
to  less than -3 for {\it all} $i$.  It can be easily seen that $c_i=-4$ is also ruled out. 
As follows from eq.(\ref{lpri}), all the $\l'_{ijk}$ except $\l'_{i11}$ are zero
in this case to start with. But the mixing of superfields in kinetic
terms can regenerate other $\l'_{ijk}$. Specifically, one gets
\beqa
\label{kmix1}
\l'_{i12} &=& V^{D}_{12}\l'_{i11} \sim \l \nonumber \\
\l'_{i21}& =& V^{Q}_{12}\l'_{i11} \sim \l  \nonumber \\
\l'_{i12} \l'_{i21} &\sim& \l^2
\eeqa
where $V^{\psi}$ rotates the matter field $\Psi_i$ to bring kinetic
terms to canonical form \cite{ramond2}
\beqa
\label{kmix2}
\Psi_i &\rightarrow& V_{ij}^{\psi} \Psi_j \nonumber \\
V_{ij}^{\psi} &\sim& \left( {< \theta > \over M} \right)^{\mid \psi_i - \psi_j \mid}
\eeqa

It follows from the above that one must require $c_i<-4$ for all $i$.
 One concludes from eq.(\ref{lpri}) that 
only phenomenologically viable possibility in model {\rm IA} is to
require vanishing $\l'_{ijk}$ for all values of $i,j,k$. We emphasize that a 
non-trivial role is played in the above argument by 
the requirement of zero or negative
$l_i+h_2$ and by the value of $h$  determined from the anomaly constraints.

The above argument also serves to restrict the trilinear 
couplings $\l_{ijk}$.  Defining the 
antisymmetric matrices $(\Lambda_k)_{ij}\equiv \lambda_{ijk}$, 
one could rewrite the $\Lambda_k$ as follows:
\beqa
(\Lambda_1)_{ij}& =&  \l^4  \left( 
\ba{ccc}
0 & \l^{c_2} & \l^{c_3} \\
-\l^{c_2} & 0 & \l^{c_3 + l_2 - l_1}\\
-\l^{c_3} & -\l^{c_3 + l_2 - l_1} & 0 \ea
\right) \nonumber \\
(\Lambda_2)_{ij}& =&  \l^2  \left( 
\ba{ccc}
0 & \l^{c_1} & \l^{c_3 + l_1 - l_2 } \\
-\l^{c_1} & 0  & \l^{c_3} \\
-\l^{c_3 +l_1 - l_2} & -\l^{c_3} & 0  \ea
\right) \nonumber \\
(\Lambda_3)_{ij}& =&   \left( 
\ba{ccc}
0 & \l^{c_2 + l_1 - l_3 } & \l^{c_1} \\
-\l^{c_2 + l_1 - l_3 } & 0  & \l^{c_2} \\
-\l^{c_1} & -\l^{c_2} & 0 \ea
\right) 
\eeqa
where $c_i$ are the same coefficients defined in the context of the $\lambda'$
and are required to be $<-4$ as argued above. It
 then immediately follows from Table 1
that all the $\l_{ijk}$ except $\l_{123},\l_{231}$ and $\l_{312}$ are forced to 
be zero. Moreover, $\l_{312}$ and $\l_{231}$ cannot simultaneously be zero.
Thus one reaches an important conclusion that Model {\rm IA} 
can be consistent with 
phenomenology only if all $\lambda'_{ijk}$ and all $\lambda_{ijk}$ except 
at most two are zero. We have not made use of 
one of the anomaly equation namely,
$A_1'=0$. Use of this does not allow even one $\l_{ijk}$ to be non-zero
in large number of models.
 
Essentially the same argument can be repeated also in case of other models.
The structure of the $\l'_{ijk}$ is determined in these models by
\beq
\l'_{ijk}\sim \lambda^{c_i+q_{j3}+d_{k3}} \eeq
where $c_i\equiv l_i+h_2+x-h$;
The main difference compared to earlier model is that the $h$ appearing in
$c_i$ is not forced to be zero but is given by eq.(\ref{h}) and can take values 
-1 ( Model {\rm IB}, Model {\rm IIIA}, Model {\rm IVB} ) or -2 ( Model {\rm IIIB} ). 
The $h=0$ for model {\rm IIB} and the above argument made in the case
of model {\rm IA} also remains valid in this case. Because, $h \leq 0$
in these models, they allow somewhat larger values for $c_i$ compared 
to $c_i \leq 2$ in case of model {\rm IA}. These larger values 
of $c_i$ result in extreme case corresponding to $l_i + h_2 = 0$ and
$x = 2$. It is possible to satisfy constraint coming from $\Delta m_K$ 
in these extreme cases e.g for model {\rm IB} $l_i + h_2 =0, x=2$ leads
\footnote{similar marginal cases are also found for models,
  {\rm IIIB,IVB}.} to 

\beq
\label{marginal}
(\l'_{i})_{jk} \approx \left (
   \begin{array}{ccc}
       \l^{7} & \l^{6} & \l^{6} \\
       \l^{6} & \l^{5} & \l^{5 }\\
       \l^{4} & \l^{3} & \l^{3} 
     \end{array}
     \right ).
\eeq

This structure is consistent with eq.(\ref{k0k0}) as well as all other
constraints on $\l'_{ijk}$. This possibility  cannot be therefore 
ruled out purely on phenomenological grounds. But as we will show, 
$A'_1 =0$ plays an important role and does not allow these marginal
cases. 

\section{Models}
Let us now discuss specific models which successfully meet all the 
phenomenological constraints. An important role is played in categorizing
these models by the anomaly constraint $A_1'=0$  which has been not yet imposed. 
Imposition of this further constraints the model. 

It is possible to give a general solution of all the anomaly constraints for all
the models listed in Table 1. We outline the solution for $A'_1 = 0$ condition 
in the appendix.  We have numerically looked for integer solutions of 
the anomaly constraints satisfying the 
criteria (1) $l_i+h_2 \leq 0$ (2) $c_i$ are chosen to 
satisfy the constraint eq.(\ref{k0k0}) e.g.
$c_i<-4$ in case of Model {\rm IA} (3) The absolute values 
of $q_3,u_3,d_3,l_1,l_2,l_3$ are restricted 
to be less than or equal to 10. The last requirement is imposed for simplicity.
Moreover in practice, higher values of these charges will generically result in 
suppressed $R$ violating couplings which may not be of phenomenological interest.
Although, all the \u1x couplings can be specified using only four parameters, we 
have displayed values of $x,q_3,u_3,d_3,l_i$ and $l_i+h_2$ in tables {\bf 2A - 2G }.
We draw the following conclusions from the tables:\\

\noindent
(1) None of the models displayed allow the value $l_i+h_2=0$ ruling out the 
marginal models displayed in eqs.(\ref{marginal}) at least for the ranges of parameters
considered here.\\

\noindent
(2) While all the $\l'_{ijk}$ are forced to be zero, some of the models allow
one or two non-zero $\l_{ijk}$. We have shown this in the last column 
which also gives the order of magnitude for the allowed $\l_{ijk}$.
 This need not always be compatible 
with phenomenology particularly after taking care of the mixing of kinetic 
energy terms. Thus some of the models displayed in tables would not be allowed.\\

\noindent
(3) Although the term $L_iH_2$ is not directly allowed, 
it can be generated from the Kahler potential through the 
mechanism proposed by GM \cite{GM} in order to 
explain the $\m$ parameter. The order of magnitudes
 of the $\e_i$ is given in this case by
\beq \e_i\sim m_{3/2}\l^{|l_i+h_2|}, \eeq

\noindent
where $m_{3/2}$ is the gravitino mass. 
This can be read off from the table in all the cases.
 Uniformly large magnitudes
of $l_i+h_2$ found in tables implies that the $R$ violation through 
effective bilinear term is also 
quite suppressed but it can still be of phenomenological relevance.\\

\noindent
(4) We did  not impose  baryon parity in the above analysis. The look at the solutions
presented in the table however shows that the operator
$U^c_iD^c_jD^c_k$ carries large negative charge  in all the models.
Thus baryon number violating terms are automatically forbidden from the 
superpotential. These terms will be generated from the effective \u1x violating 
$D$ term
$$ {1\over M_P}({\theta^*\over M})^{\mid q_{ijk} \mid} (U^c_iD^c_jD^c_k)$$ 
where $q_{ijk}$ is the negative \u1x charge of 
the combination $U^c_iD^c_jD^c_k$.
This leads to baryon number violating couplings 
$$\l''_{ijk}\sim {m_{3/2}\over M_P} \l^{|q_{ijk}|}$$
which are extremely suppressed, $\leq O(10^{-15})$ for $m_{3/2} \sim $ TeV.
Thus proton stability gets 
automatically explained in all the models.\\

\noindent
(5) The trilinear lepton number violating terms are  not allowed 
in the superpotential from analyticity. But they  will be effectively
 generated in the same way as $\l''$ discussed above.  Their 
magnitudes will also be enormously suppressed $\leq 10^{-15}$ depending 
upon the model. 

It follows from the forgoing discussions that consistently 
implemented \u1x symmetry
allows very simple $R$ violating interactions namely three bilinear terms and
at most two  trilinear coupling $\l_{ijk}$. The constraints 
coming from the $K^0-\bar{K^0}$  mass difference were instrumental in 
arriving at this conclusion.  It is worth emphasizing that the effective bilinear
 interactions  generated from GM mechanism in this case are not subject to
such stringent constraint from the flavour violating process. A priori, 
the bilinear terms can be rotated away in favour of trilinear $\l'$ and $\l$ 
interactions. It turns out that one does not generate dangerous flavour 
violating terms in the process. Specifically, one 
finds for the flavour structure \cite{asjbabu},
\beq 
\label{rottri}
W= -{\tan\theta_3\over <H_1>}[(O_L^T)_{3\a}L_{\a}]\left(
 m^l_{\b}L_{\b}e^c_{\b}+m_i^DQ_id_i^c \right).
\eeq 
where all the fields are in the physical i.e, the mass basis. $(O_L^T)$ 
represents a mixing matrix determined solely by the ratios of $\e_i$
and  $\tan\theta_3 = \sqrt(\sum_i \e_i^2)/\mu  $ and $\a, \b$ run over $e,\m,\tau$.
It is seen that the 
resulting trilinear interactions are flavour diagonal and thus the parameter
$\e_i$ are not severely constrained \footnote{The same conclusion was also drawn
in ref. \cite{dudas1} by using different leptonic basis.}. The major effect of the bilinear terms
is to generate the neutrino masses and leptonic Kobayashi Maskawa matrix.

The neutrino masses in the presence of bilinear terms alone, have been discussed
in many papers \cite{bilinears}. Large number of these concentrated on universal 
boundary conditions since they provide natural means to 
understand smallness of neutrino masses even when 
the bilinear parameters are not suppressed \cite{bilinears,asjbabu}. 
The soft SUSY breaking terms are also subject
to the \u1x symmetry and need not follow the universal structure \cite{dudas3}.
 But the smallness of neutrino masses
follows here from the \u1x symmetry itself without invoking universal boundary 
conditions since the allowed values of $|l_i+h_2|$ in various tables are large
leading to suppressed ${\e\over \m}$ and hence neutrino masses, eq.( \ref{numass}). 
The detailed structure of neutrino masses and mixing will be more model 
dependent here than in
case of the universal boundary conditions. It seems possible to
 obtain reasonable mixing and masses in some of the models.
As an example, 
consider model 2 in table {\bf 2 A}. This is characterized by three bilinear terms
of equal magnitudes. Thus in the absence of any fine tuning one can expect to
get large mixing angles naturally. The heaviest neutrino would have mass of the order
$$ m_\n\sim \l^{18}{M_Z^2\over M_{SUSY}}\sim 10^{-1} \eV$$
which is in the right range for solving the atmospheric neutrino anomaly.
The other mass gets generated radiatively through eq.(\ref{rottri}) and would be suppressed
compared to the above mass. The detailed predictions of the neutrino spectrum would 
depend upon the structures of soft symmetry breaking terms which themselves
would be determined by the \u1x symmetry. We shall not discuss it here.
\section{Summary}
The supersymmetric standard model allows 39 lepton number violating parameters
which are not constrained theoretically. We have shown in this paper that the 
\u1x symmetry invoked to understand fermion masses can play an important role 
in constraining these parameters. We restricted ourselves to integer \u1x charges and 
considered  different \u1x charge assignments compatible with fermion spectrum.
We have shown that  only phenomenologically consistent possibility in this context 
is that all the trilinear $\l'_{ijk}$ and all but two  $\l_{ijk}$ couplings to be zero or
extremely small of $O(10^{-15})$. While the patterns of $R$ violation have been earlier 
discussed in the presence of \u1x symmetry the systematic confrontation of these
pattern with phenomenology leading to this important conclusion was not made to 
the best of our knowledge.
In fact, some works \cite{rpreview} 
which neglected important constraint of $l_i + h_2 \leq 0$ 
concluded to the contrary that it is possible to obtain phenomenologically consistent
and non-zero trilinear couplings.

Our work is restricted to only \u1x symmetry which is by far  most popular and to 
integer \u1x charges.  Use of other horizontal symmetries can allow non-zero trilinear 
interactions and still
be consistent with phenomenology. An example of this can be found in \cite{nir}.
Our work is closely related to and compliments the analysis presented in \cite{dudas2}.
It was assumed in this paper that bilinear $R$ violating interactions come from
the GM mechanism and are absent in the superpotential. Assuming that there are no
trilinear interactions in the superpotential it was shown that flavor violating 
transitions in the model are adequately suppressed. We have systematically shown that 
this is the only allowed possibility except for the occurrence of one or two trilinear
$\l_{ijk}$ couplings. This way, \u1x symmetry is shown to require that only
four or five of the total 39 lepton number violating couplings could have magnitudes
in the phenomenologically interesting range! 

\section{Appendix}
Here we give the most general solutions for the Green-Schwarz anomaly 
conditions in terms of the four independent charges. The constraints
 ~$A_3 = A_2~ \mbox{and}~ A_3 = {3 \over 5} ~A_1 $ gave us eq.(\ref{h}).
The condition $A'_1 =0$ can be solved to give,
\beq
l_3 = A ~d_3 + B~ q_3 + C ~l_1 + D~ x + E 
\eeq
where
\beqa
A&=& {-1 \over k_2} \left( \sum_i (d_{i3} + 2 u_{i3}) - h + k_1 
+ k_2 - m + 3 x \right) \nonumber \\
B&=& {-1 \over k_2} \left( \sum_i (q_{i3} + 4 u_{i3}) - 7 h + k_1 
+ 10 k_2 - m + 9 x \right) \nonumber \\
C&=& {-1 \over k_2} \left( k_2 - k_1 \right) \nonumber \\
D&=& {-1 \over k_2} \left( 5 h  - 4 \sum_i (u_{i3})
  - 3 (k_2 + x) \right) \nonumber \\
E&=& \left( \sum_i (d_{i3}^2 +  q_{i3}^2  -2 u_{i3}^2 + k_i^2 ) - 5 h^2 
+ 2 k_2 (4 h - m ) \right)
\eeqa
and
\beqa
k_1 &=& l_{13} + e_{13} \nonumber \\
k_2 &=& l_{23} + e_{23} 
\eeqa
In the above we have taken $q_3,d_3,l_1$ and $x$ as four independent parameters
and $l_3$ has been expressed in terms of them. $m$ and $u_3$ are respectively
 given by eqs.(\ref{em},\ref{u3}) 
of the text and remaining charges by the Table 1 defining the models. This way all
the \u1x charges get fixed in terms of $q_3, d_3,l_1 \mbox{and}~ x$ once a model 
displayed in the table is chosen.

\newpage

\begin{center}
{\large Model IA}\\[20pt]
\begin{tabular}{|c|c|c|c|c|c|c|c|c|c|c|c|} \hline
{\bf No.} & {\bf x} & {$\bf q_3$} & {$\bf u_3$} & {$\bf d_3$} & {$\bf l_1$} & {$\bf l_2$} & {$\bf l_3$}
& {$\bf f_1$} & {$\bf f_2$} & {$\bf f_3$} & {\bf If $\bf \lambda_{ijk}$
allowed} \\ \hline\hline

1 & 0 & 2 & 1 & -5 & -6 & -3 & -6 & -9 & -6 & -9 & No \\ 
2 & 0 & 2 & 1 & -5 & -5 & -5 & -5 & -8 & -8 & -8 & No \\ 
3 & 0 & 2 & 1 & -5 & -4 & -7 & -4 & -7 & -10 & -7 & No \\ 
4 & 0 & 2 & 1 & -5 & -3 & -9 & -3 & -6 & -12 & -6 & $\lambda_{132} \sim
4.8 \times 10^{-2} $ \\ 
5 & 0 & 2 & 2 & -6 & -10 & -4 & -1 & -14 & -8 & -5 & $\lambda_{231} \sim 
5.1 \times 10^{-4}$ \\ 
6 & 0 & 3 & 2 & -8 & -10 & -4 & -10 & -15 & -9 & -15 & No \\ 
7 & 0 & 3 & 2 & -8 & -9 & -6 & -9 & -14 & -11 & -14 & No \\ 
8 & 0 & 3 & 2 & -8 & -8 & -8 & -8 & -13 & -13 & -13 & No \\ 
9 & 0 & 3 & 2 & -8 & -7 & -10 & -7 & -12 & -15 & -12 & No \\ 
10 & 2 & 3 & 2 & -6 & -7 & -3 & -8 & -12 & -8 & -13 & No \\ 
11 & 2 & 3 & 2 & -6 & -6 & -5 & -7 & -11 & -10 & -12 & No \\ 
12 & 2 & 3 & 2 & -6 & -5 & -7 & -6 & -10 & -12 & -11 & No \\ 
13 & 2 & 3 & 2 & -6 & -4 & -9 & -5 & -9 & -14 & -10 & No \\ 
14 & 2 & 4 & 3 & -9 & -9 & -8 & -10 & -16 & -15 & -17 & No \\ 
15 & 2 & 4 & 3 & -9 & -8 & -10 & -9 & -15 & -17 & -16 & No \\ \hline
 
\end{tabular}
\end{center}
{\bf Table 2A}: Here we display the allowed models where the following constraints have
been imposed : a) requirement of correct quark and lepton mass hierarchies as per Model {\rm IA}
in table I b) GS anomaly cancellations c) $f_i = l_i+h_2 \leq 0 $ d) phenomenological
constraints from $K^0-{\bar K}^0$ mixing on $\l'_{ijk}$ couplings  and 
(e) $|q_3,u_3,d_3,l_i| \leq$ 10.

\newpage

\begin{center}
{\large Model IB}\\[20pt]
\begin{tabular}{|c|c|c|c|c|c|c|c|c|c|c|c|} \hline
{\bf No.} & {\bf x} & {$\bf q_3$} & {$\bf u_3$} & {$\bf d_3$} & {$\bf l_1$} & {$\bf l_2$} & {$\bf l_3$}
& {$\bf f_1$} & {$\bf f_2$} & {$\bf f_3$} & {\bf If $\bf \lambda_{ijk}$
allowed} 
\\ \hline\hline

1 & 0 & 2 & 2 & -5 & -6 & -3 & -2 & -10 & -7 & -6 & $\lambda_{131} \sim
1.0, \lambda_{231} \sim 10^{-2}$ \\ 
2 & 0 & 3 & 2 & -7 & -4 & -6 & -10 & -9 & -11 & -15 & No \\ 
3 & 1 & 3 & 2 & -6 & -3 & -5 & -9 & -8 & -10 & -14 & No \\ 
4 & 0 & 3 & 3 & -8 & -10 & -1 & -9 & -16 & -7 & -15 & $\lambda_{231} \sim   
1.0 $ \\
5 & 0 & 3 & 3 & -8 & -8 & -6 & -6 & -14 & -12 & -12 & No \\ 
6 & 1 & 3 & 3 & -7 & -8 & -4 & -5 & -14 & -10 & -11 & $\lambda_{231} \sim 
1.0 $ \\ 
7 & 1 & 3 & 3 & -7 & -6 & -9 & -2 & -12 & -15 & -8 & No \\ 
8 & 2 & 3 & 3 & -6 & -8 & -2 & -4 & -14 & -8 & -10 & $\lambda_{121} \sim 
1.0, \lambda_{231} \sim 2.3 \times 10^{-3} $ \\ 
9 & 1 & 4 & 4 & -10 & -10 & -7 & -9 & -18 & -15 & -17 & No \\ 
10 & 2 & 4 & 4 & -9 & -10 & -5 & -8 & -18 & -13 & -16 & No \\ 
11 & 2 & 4 & 4 & -9 & -8 & -10 & -5 & -16 & -18 & -13 & No \\  \hline

\end{tabular}
\end{center}
{\bf Table 2B}: Same as above, but for values given by Model {\rm IB}.
%\newpage
\vskip 1 cm
\begin{center}
{\large Model IIB}\\[20pt]
\begin{tabular}{|c|c|c|c|c|c|c|c|c|c|c|c|} \hline
{\bf No.} & {\bf x} & {$\bf q_3$} & {$\bf u_3$} & {$\bf d_3$} & {$\bf l_1$} & {$\bf l_2$} & {$\bf l_3$}
& {$\bf f_1$} & {$\bf f_2$} & {$\bf f_3$} & {\bf If $\bf \lambda_{ijk}$
allowed} \\
\hline\hline

1 & 0 & 2 & 2 & -6 & -3 & -8 & -9 & -7 & -12 & -13 & No \\ 
2 & 0 & 2 & 3 & -7 & -8 & -5 & -7 & -13 & -10 & -12 & No \\ 
3 & 0 & 2 & 3 & -7 & -6 & -10 & -4 & -11 & -15 & -9 & No \\ 
4 & 1 & 2 & 3 & -6 & -8 & -2 & -7 & -13 & -7 & -12 & $\lambda_{231} \sim 
1.0 $\\ 
5 & 1 & 2 & 3 & -6 & -6 & -7 & -4 & -11 & -12 & -9 & No \\ 
6 & 2 & 2 & 3 & -5 & -6 & -4 & -4 & -11 & -9 & -9 & $\lambda_{231} \sim  
1.0$ \\
7 & 1 & 3 & 4 & -9 & -9 & -10 & -7 & -16 & -17 & -14 & No \\ 
8 & 2 & 3 & 4 & -8 & -9 & -7 & -7 & -16 & -14 & -14 & No \\ \hline

\end{tabular}
\end{center}

{\bf Table 2C}: Same as above, but for values given by Model {\rm IIB}.
\newpage

\begin{center}
{\large Model IIIA}\\[20pt]
\begin{tabular}{|c|c|c|c|c|c|c|c|c|c|c|c|} \hline
{\bf No.} & {\bf x} & {$\bf q_3$} & {$\bf u_3$} & {$\bf d_3$} & {$\bf l_1$} & {$\bf l_2$} & {$\bf l_3$}
& {$\bf f_1$} & {$\bf f_2$} & {$\bf f_3$} & {\bf If $\bf \lambda_{ijk}$
allowed} \\ \hline\hline

1 & 0 & 2 & 3 & -6 & -7 & -2 & -9 & -12 & -7 & -14 & No \\ 
2 & 0 & 2 & 3 & -6 & -6 & -4 & -8 & -11 & -9 & -13 & No \\ 
3 & 0 & 2 & 3 & -6 & -5 & -6 & -7 & -10 & -11 & -12 & No \\ 
4 & 0 & 2 & 3 & -6 & -4 & -8 & -6 & -9 & -13 & -11 & No \\ 
5 & 0 & 2 & 3 & -6 & -3 & -10 & -5 & -8 & -15 & -10 & $\lambda_{132} \sim
1.0$ \\ 
6 & 1 & 2 & 3 & -5 & -6 & -2 & -7 & -11 & -7 & -12 & No \\ 
7 & 1 & 2 & 3 & -5 & -5 & -4 & -6 & -10 & -9 & -11 & No \\ 
8 & 1 & 2 & 3 & -5 & -4 & -6 & -5 & -9 & -11 & -10 & No \\ 
9 & 1 & 2 & 3 & -5 & -3 & -8 & -4 & -18 & -13 & -9 & $\lambda_{132} \sim
1.0$ \\ 
10 & 1 & 2 & 3 & -5 & -2 & -10 & -3 & -7 & -15 & -8 & $\lambda_{132} \sim 
2.3 \times 10^{-3}$\\ 
11 & 2 & 2 & 3 & -4 & -4 & -4 & -4 & -9 & -9 & -9 & No \\ 
12 & 2 & 2 & 3 & -4 & -3 & -6 & -3 & -8 & -11 & -8 & $\lambda_{132} \sim 
1.0$ \\ \hline

\end{tabular}
\end{center}
{\bf Table 2D}: Same as above, but for values given by Model {\rm IIIA}.
%\newpage
\vskip 1 cm

\begin{center}
{\large Model IIIB}\\[20pt]
\begin{tabular}{|c|c|c|c|c|c|c|c|c|c|c|c|} \hline
{\bf No.} & {\bf x} & {$\bf q_3$} & {$\bf u_3$} & {$\bf d_3$} & {$\bf l_1$} & {$\bf l_2$} & {$\bf l_3$}
& {$\bf f_1$} & {$\bf f_2$} & {$\bf f_3$} & {\bf If $\bf \lambda_{ijk}$
allowed} \\ \hline\hline

1 & 0 & 2 & 3 & -5 & -2 & -5 & -7 & -7 & -10 & -12 & No \\ 
2 & 0 & 2 & 4 & -6 & -7 & -3 & -4 & -13 & -9 & -10 & $\lambda_{231} \sim 
0.22$ \\ 
3 & 0 & 2 & 4 & -6 & -5 & -8 & -1 & -11 & -14 & -7 & $\lambda_{131} \sim 
1.0, \lambda_{132} \sim 1.0$ \\ 
4 & 2 & 3 & 4 & -6 & -3 & -4 & -10 & -10 & -11 & -17 & $\lambda_{123}
\sim 1.0 $\\ 
5 & 0 & 3 & 5 & -9 & -8 & -6 & -9 & -16 & -14 & -17 & No \\ 
6 & 1 & 3 & 5 & -8 & -9 & -3 & -8 & -17 & -11 & -16 & No \\ 
7 & 1 & 3 & 5 & -8 & -7 & -8 & -5 & -15 & -16 & -13 & No \\ 
8 & 2 & 3 & 5 & -7 & -8 & -5 & -4 & -16 & -13 & -12 & $\lambda_{231} \sim
1.0$ \\ 
9 & 2 & 3 & 5 & -7 & -6 & -10 & -1 & -14 & -18 & -9 & $\lambda_{131} \sim
1.0, \lambda_{132} \sim 0.22$ \\ 
10 & 2 & 4 & 6 & -10 & -9 & -8 & -9 & -19 & -18 & -19 & No \\  \hline

\end{tabular}
\end{center}

{\bf Table 2E}: Same as above, but for values given by Model {\rm IIIB}.

\newpage

\begin{center}
{\large Model IVA}\\[20pt]
\begin{tabular}{|c|c|c|c|c|c|c|c|c|c|c|c|} \hline
{\bf No.} & {\bf x} & {$\bf q_3$} & {$\bf u_3$} & {$\bf d_3$} & {$\bf l_1$} & {$\bf l_2$} & {$\bf l_3$}
& {$\bf f_1$} & {$\bf f_2$} & {$\bf f_3$} & {\bf If $\bf \lambda_{ijk}$
allowed} \\ \hline\hline

1 & 0 & 6 & -3 & -9 & -10 & -4 & -7 & -13 & -7 & -10 & $\lambda_{231} \sim
1.0$ \\ 
2 & 0 & 6 & -3 & -9 & -9 & -6 & -6 & -12 & -9 & -9 & No \\ 
3 & 0 & 6 & -3 & -9 & -8 & -8 & -5 & -11 & -11 & -8 & No \\ 
4 & 0 & 6 & -3 & -9 & -7 & -10 & -4 & -10 & -13 & -7 & No \\ 
5 & 2 & 7 & -2 & -10 & -8 & -6 & -10 & -13 & -11 & -15 & No\\ 
6 & 2 & 7 & -2 & -10 & -7 & -8 & -9 & -12 & -13 & -14 & No\\ 
7 & 2 & 7 & -2 & -10 & -6 & -10 & -8 & -11 & -15 & -13 & No\\ \hline

\end{tabular}
\end{center}

{\bf Table 2F}: Same as above, but for values given by Model {\rm IVA}.

\vskip 1 cm
\begin{center}
{\large Model IVB}\\[20pt]
\begin{tabular}{|c|c|c|c|c|c|c|c|c|c|c|c|} \hline
{\bf No.} & {\bf x} & {$\bf q_3$} & {$\bf u_3$} & {$\bf d_3$} & {$\bf l_1$} & {$\bf l_2$} & {$\bf l_3$}
& {$\bf f_1$} & {$\bf f_2$} & {$\bf f_3$} & {\bf If $\bf{\lambda_{ijk}}$
allowed }\\ \hline\hline

1 & 0 & 6 & -2 & -9 & -8 & -5 & -4 & -12 & -9 & -8 & $\lambda_{231} \sim
0.22$ \\ 
2 & 2 & 7 & -1 & -10 & -8 & -5 & -7 & -14 & -11 & -13 & No \\ 
3 & 2 & 7 & -1 & -10 & -6 & -10 & -4 & -12 & -16 & -10 & No \\ \hline

\end{tabular}
\end{center}

{\bf Table 2G}: Same as above, but for values given by  Model {\rm IVB}.

\end{document}